\documentclass[12pt]{iopart}

\usepackage{epsfig,amssymb}

\def\spose#1{\hbox to 0pt{#1\hss}}
\def\lesssim{\mathrel{\spose{\lower 3pt\hbox{$\mathchar"218$}}
 \raise 2.0pt\hbox{$\mathchar"13C$}}}
\def\gtrsim{\mathrel{\spose{\lower 3pt\hbox{$\mathchar"218$}}
 \raise 2.0pt\hbox{$\mathchar"13E$}}}

\begin{document}
\title{
Critical behavior of the random-anisotropy model in the strong-anisotropy 
limit }
\author{
Francesco Parisen Toldin,$^{1,3}$ 
Andrea Pelissetto,$^2$ and Ettore Vicari$^3$ 
}
\address{$^1$ Scuola Normale Superiore, Pisa, Italy
}
\address{$^2$ Dipartimento di Fisica dell'Universit\`a di Roma 
``La Sapienza" and INFN, Roma, Italy.}
\address{$^3$
Dipartimento di Fisica dell'Universit\`a di Pisa 
and INFN, Pisa, Italy.
}
\ead{
{\tt f.parisentoldin@sns.it},
{\tt Andrea.Pelissetto@roma1.infn.it},
{\tt Ettore.Vicari@df.unipi.it}
}

\date{\today}

\maketitle

\begin{abstract}
We investigate the nature of the critical behavior of the random-anisotropy
    Heisenberg model (RAM), which describes a magnetic system with random
    uniaxial single-site anisotropy, such as some amorphous alloys of
    rare earths and transition metals. In particular, we consider the
    strong-anisotropy limit (SRAM), in which the Hamiltonian can be
    rewritten as the one of an Ising spin-glass model with correlated bond
    disorder.  We perform Monte Carlo
    simulations of the SRAM on simple cubic lattices of linear size 
    $L$, up to $L=30$,
    measuring correlation functions of the replica-replica overlap,
    which is the order parameter at a glass transition. The corresponding
    results show critical behavior and finite-size scaling. They provide
    evidence of a finite-temperature continuous transition with critical
    exponents $\eta_o=-0.24(4)$ and $\nu_o=2.4(6)$. These results are
    close to the corresponding estimates that have been obtained in
    the usual Ising spin-glass model with uncorrelated  bond disorder,
    suggesting that the two models belong to the same universality class.
    We also determine the leading correction-to-scaling exponent, finding 
    $\omega = 1.0(4)$.
\end{abstract}

\pacs{75.50.Lk, 05.70.Jk, 75.40.Mg, 77.80.Bh}


\section{Introduction}
\label{secint}

The critical behavior of magnetic systems in the presence of quenched disorder
has been the subject of extensive theoretical and experimental study.  An
important class of systems consists in amorphous alloys of rare earths with
aspherical electron distributions and transition metals, for instance TbFe$_2$
and YFe$_2$. They are modeled \cite{HPZ-73} by a Heisenberg model
with random uniaxial single-site anisotropy defined on a simple cubic lattice,
or, in short, by the random-anisotropy model (RAM)
\begin{equation}
{\cal H} =- J \sum_{\langle xy \rangle} \vec{s}_x \cdot \vec{s}_y-
D \sum_x(\vec{u}_x \cdot \vec{s}_x)^2,
\label{HRAM}
\end{equation}
where $\vec{s}_x$ is a three-component spin variable, $\vec{u}_x$ is a unit
vector describing the local (spatially uncorrelated) random anisotropy, and
$D$ the anisotropy strength. In amorphous alloys the distribution of
$\vec{u}_x$ is usually taken to be isotropic, since, in the absence of
crystalline order, there is no preferred direction.  

Random anisotropy is a relevant perturbation of the pure Heisenberg model, 
so that random-anisotropy systems show a critical behavior that is different 
from the Heisenberg one. For small $D$ however, Heisenberg behavior may still 
be observed, the Heisenberg fixed point controlling the multicritical behavior
of the system.  If $T_p\equiv T_c(D=0)$ is the critical temperature of the 
pure Heisenberg model,
in the limit $t_p\equiv T/T_p-1 \to 0$ and $D\to 0$ 
the free energy has the scaling form \cite{CPV-04}
\begin{equation}
{\cal F}= |t_p|^{2-\alpha_H}f(D^2 |t_p|^{-\phi_D}),
\label{freeen}
\end{equation}
where $\alpha_H=-0.1336(15)$ is the specific-heat exponent in the pure
Heisenberg theory \cite{CHPRV-02,PV-rev}, $\phi_D=0.412(3)$ is the
crossover exponent associated with the random-anisotropy perturbation
\cite{CPV-04}, and $f(x)$ is a universal function.\footnote{
As a consequence of Eq.~(\ref{freeen}), for sufficiently small $D$ the
critical-temperature shift is given by $T_c(D)- T_c(0) \approx c
D^{2/\phi_D} + a_1 D^2 + a_2 D^4 + \ldots$ where $2/\phi_D\approx
4.9$.  Note that the nonanalytic term $D^{2/\phi_D}$ is suppressed
with respect to the first two analytic terms $D^2$ and $D^4$.}

The critical behavior in the presence of
random anisotropy has been
investigated at length, but a satisfactory picture has not
been achieved yet. A recent review can be found in
\cite{DFH-05}. The Imry-Ma argument \cite{IM-73,PPR-78} forbids
the appearance of a low-temperature phase with nonvanishing
magnetization for $d<4$.  This is also supported by field-theoretical
renormalization-group (RG) calculations using the replica
method~\cite{Aharony-75,MG-82,DFH-01a,CPV-04}. However, 
a glassy transition with a low-temperature phase characterized 
by quasi-long-range order (QLRO), i.e., a phase 
in which correlation functions decay
algebraically, is still possible \cite{PPR-78}. A Landau-Ginzburg calculation 
\cite{AP-80} of the equation of state for $D\to 0$ 
and a recent $4-\epsilon$ study \cite{Feldman-00,Feldman-01} based on a
functional RG approach support this scenario in the weak-anisotropy limit.
In the large-anisotropy limit $D\to \infty$ the model becomes an 
Ising spin glass with a correlated bond distribution. 
An interesting hypothesis, originally put
forward in \cite{CL-77}, is that in this limit the RAM
transition is in the same universality class as that of 
the Ising spin-glass model (ISGM) \cite{MPV-87,FH-91,KR-03}.

Numerical simulations (see, for example,
\cite{JK-80,Chakrabarti-87,Fisch-90,Fisch-98,Itakura-03})
provide some evidence of the existence of a finite-temperature
transition for small values of $D$.  On the other hand, for large $D$, 
even the presence
of a finite-temperature transition is not supported by numerical
simulations \cite{Itakura-03}.
Experiments support the absence of a
ferromagnetic phase for amorphous systems in general, and provide
evidence of a generic glassy behavior at sufficiently low
temperature.  The nature of the transition and of the low-temperature
phase is however unclear. In particular, no evidence of QLRO has been
reported.

The above-reported arguments apply to the RAM in which 
anisotropy has a generic cubic-symmetric distribution.
However, in some particular cases, when disorder preserves
the reflection symmetry $s_{x,a}\to -s_{x,a}$, $s_{x,b}\to s_{x,b}$ 
for $b\not= a$ (this is,
for example, realized when the probability distribution vanishes outside the
lattice axes), there is a standard order-disorder transition with a
low-temperature magnetized phase.  Moreover, a RG analysis of the
corresponding Landau-Ginzburg-Wilson theory~\cite{CPV-04,MG-82,HBDFFY-03}
shows that continuous transitions belong to the same universality class 
as that of the
random-exchange Ising model (REIM).\footnote{The main difference 
with respect to the REIM critical
behavior (see, e.g., \cite{PV-rev,FHY-03}) is the approach to the
asymptotic critical behavior, which is controlled by scaling corrections
with exponent $\Delta = - \alpha_r$, where $\alpha_r\simeq -0.05$ is the
specific-heat exponent of the REIM \cite{CPV-04}.
This is much smaller than the correction-to-scaling 
exponent of the REIM, which is $\Delta_{\rm REIM}\approx 0.25$.}

In this paper we investigate the critical behavior of the RAM
for a uniform distribution in the limit $D/J\rightarrow \infty$.
In this case
we can write $\vec{s}_x=\sigma_x \vec{u}_x$, with $\sigma_x=\pm 1$.
Thus, the RAM reduces to a particular Ising spin-glass model with Hamiltonian
\cite{JK-80}
\begin{equation}
{\cal H} = - \sum_{\langle xy \rangle} j_{xy} \sigma_x \sigma_y,
\qquad j_{xy}= \vec{u}_x \cdot \vec{u}_y,
\label{raminfD} 
\end{equation}
which we call strong random-anisotropy model (SRAM) (We set $J=1$ without
loss of generality).
Model (\ref{raminfD}) differs from the usual ISGM in the bond distribution.
Here the random variables $j_{xy}$ on different lattice links are 
correlated. For instance, one has 
$\overline{\prod_\square j_{xy}} = 1/27$, 
where the product is over the links belonging to a given 
plaquette and the average is taken with respect
to the distribution of the vectors $\vec{u}_x$.
Thus, the SRAM is not only of interest as a model of a class of magnetic 
systems, but also {\em per se}, to understand how glassy behavior 
in Ising systems depends on the disorder distribution.

We study the critical behavior of the SRAM 
by means of Monte Carlo (MC) simulations. Since the model is 
essentially a spin glass we shall focus on the so-called overlap 
variables \cite{MPV-87} $\sigma_x\tau_x$, where $\sigma_x$ and $\tau_x$
are associated with two different replicas of the model with the same 
bond variables. For the SRAM one can also consider the 
standard magnetic variables $\vec{s}_x = \sigma_x \vec{u}_x$.
The MC results of \cite{Itakura-03} suggest that these
quantities are not critical. We confirm here these conclusions so that 
little will be said about magnetic variables in the 
present study. We will study the behavior of the SRAM in the high-temperature 
phase. This reduces the algorithmic problems---the MC algorithm
becomes very slow as temperature is reduced---and allows us to consider 
lattices of size $L^3$ up to $L = 30$. In order to take into account 
finite-size effects we use the iterative method based 
on finite-size scaling (FSS) introduced in \cite{CEFPS-95}
and applied to disordered systems in 
\cite{PC-99,CMPV-03,Joerg-06,JLEM-06}.
It allows us to obtain infinite-volume results up to
$\xi_\infty \approx 20$ ($\xi_\infty$ is the infinite-volume 
second-moment correlation length associated with the overlap two-point
correlation function) in the high-temperature phase. 

Our MC results do not show a direct evidence of a phase transition for 
$\beta \lesssim 1.00$, which is the range of values of $\beta$ that we can
reliably simulate. On the other hand, our data clearly show FSS 
as $\beta$ increases, providing indirect evidence of the presence of 
a critical point. Fits of the infinite-volume results obtained for 
$\beta \lesssim 0.95$ indicate $\beta_c = 1.08 \pm 0.04$. The corresponding 
critical exponents are:
\begin{eqnarray}
   \eta_o =  - 0.24(4) \qquad\qquad
   \nu_o = 2.4(6).
\label{results}
\end{eqnarray}
They are defined by $\chi \sim \xi^{2- \eta_o}$ and 
$\xi \sim (\beta_c - \beta)^{-\nu_o}$; the suffix $o$ is introduced to 
remind that all exponents refer to the overlap variables and not to the 
magnetic ones. In the analysis it is crucial to include corrections
to scaling in the FSS (corrections behave as 
$L^{-\omega}$) and in fits of infinite-volume 
quantities (nonanalytic corrections behave as $\xi_\infty^{-\omega}$ as 
$\xi_\infty\to\infty$). 
The exponent
$\omega$ has been determined by studying the critical behavior of two 
universal ratios that involve the four-point and the two-point correlation
function of the overlap variables. We obtain
\begin{equation}
\omega = 1.0 \pm 0.4.
\end{equation}
Estimates (\ref{results}) are reasonably close to those obtained for 
the ISGM (see Table 1 in \cite{KKY-06} for a list of recent 
results) and thus support the conjecture that the SRAM transition 
is in the same universality class as that of the ISGM. 
Some additional arguments will be presented in Sec.~\ref{conclusions}.
Similar conclusions were reached in \cite{BM-85} for the two-dimensional
case. By using the large-cell RG method, it was shown that the 
two-dimensional SRAM has a zero-temperature transition with
critical exponents compatible with those of the 
two-dimensional ISGM.

The paper is organized as follows. In Sec.~\ref{def} we define the quantities
that will be studied numerically. In Sec.~\ref{mcsim} we 
discuss the MC algorithm. In Sec.~\ref{res} we  
discuss our MC simulations, providing evidence for the 
existence of a critical transition and computing the corresponding
critical exponents. Finally, in Sec.~\ref{conclusions} we draw our 
conclusions.

\section{Definitions}
\label{def}

We define here the quantities that have been determined in the 
MC simulation.
The energy density and specific heat are defined as 
\begin{equation}
E \equiv {1\over V} \overline{ \langle - {\cal H} \rangle }, \qquad
C \equiv {1\over V} \beta^2 \left( \overline{ \langle {\cal H}^2 \rangle -
\langle {\cal H} \rangle^2 } \right),
\label{energy}
\end{equation}
where $V\equiv L^3$ is the volume.
In our numerical work we focus on the critical behavior of the overlap parameter
\begin{equation}
q_x \equiv \sigma_x \tau_x, 
\label{ovpar}
\end{equation}
where $\sigma_x$ and $\tau_x$ are two independent replicas of the system
with the same couplings
$j_{xy}$.  We consider the correlation function 
$G(x) \equiv \overline{ \langle q_0 q_x \rangle }$, 
its Fourier transform $\widetilde{G}(p)$, 
the corresponding susceptibility $\chi$, and the second-moment
correlation length $\xi$:
\begin{eqnarray}
\chi \equiv  \sum_{x} G(x) = \widetilde{G}(0), \qquad\qquad
\xi^2 \equiv  {1\over 4 \sin^2 (p_{\rm min}/2)} 
{\widetilde{G}(0) - \widetilde{G}(p)\over \widetilde{G}(p)},
\label{xidefffxy}
\end{eqnarray}
where $p = (p_{\rm min},0,0)$, and $p_{\rm min} \equiv 2 \pi/L$.

Moreover, we consider quartic correlations of the overlap
parameter $q_x$ at zero momentum. Setting
$\mu_{k} \equiv \langle (\sum_x q_x)^k \rangle$, 
we define the quartic susceptibilities 
\begin{eqnarray}
V \chi_4 \equiv  
\overline{\mu_{4}} - 3 \overline{ \mu_{2}^2},
\qquad 
V \chi_{22} \equiv 
\overline{ \mu_{2}^2} - \overline{\mu_{2}}^{\,2}, 
\end{eqnarray}
and the quartic cumulant
\begin{equation}
B_q = { \overline{\mu_{4}} \over \overline{\mu_{2}}^{\,2} }\; .
\label{binder}
\end{equation}
Finally, we define the zero-momentum four-point couplings 
\begin{eqnarray}
G_4 \equiv - {\chi_4 \over \xi^3 \chi^2}  ,\qquad
G_{22} \equiv - {\chi_{22} \over \xi^3 \chi^2}.
\label{fourpoint-def} 
\end{eqnarray}
We shall also briefly consider magnetic variables associated
with $\vec{s}_x = \sigma_x\vec{u}_x$. In particular, if
$\nu_{k} \equiv \langle (\sum_{x,y} \vec{s}_x\cdot\vec{s}_y)^{k/2} 
\rangle$, we define the usual Binder cumulant
\begin{equation}
B_m \equiv  { \overline{\nu_{4}} \over \overline{\nu_{2}}^{\,2} }\; .
\label{binders}
\end{equation}

\section{The algorithm}
\label{mcsim}

We consider model (\ref{raminfD}) on simple cubic lattices $L^3$ with
periodic boundary conditions.  
No efficient algorithm is known for generic spin-glass systems 
(but recently progress has been made in some specific cases, see 
\cite{Houdayer-01,WS-05,Joerg-05} and references therein),
and thus we simply used the Metropolis algorithm with 
lexicographic choice of the lattice site. In order to reduce the 
thermalization times, which are the main source of bias in our 
simulations (and also the autocorrelation times, though this is not 
crucial, since the error on the results
is mainly due to sample-to-sample fluctuations)
we have combined it with the random-exchange
method---also called parallel tempering or multiple Markov chain
method---introduced in
\cite{Geyer-91,GT-95,HN-96} 
(for a recent review with many different applications
see \cite{ED-05}; some improvements of the method are discussed in
\cite{KTHT-06}).

In practice, we work as follows. We consider $N_{\beta}$ configurations
at different inverse temperatures $\beta_i$ ($\beta \equiv 1/T$), 
$i=1,\ldots N_\beta$, in a given range $[\beta_{\rm min},\beta_{\rm max}]$,
that evolve according to Hamiltonian (\ref{raminfD}) with the same 
couplings $j_{xy}$. Every $N_{\rm ex}$ iterations we perform 
$N_{\beta}-1$ random-exchange
moves, trying to swap sequentially $\beta_1$ with $\beta_2$, 
then $\beta_2$ with $\beta_3$, up to $\beta_{N_\beta-1}$ and $\beta_{N_\beta}$. 
Each time we propose a swap of adjacent temperatures 
$\beta_i$ and $\beta_{i+1}$, which is 
accepted with probability ${\rm Min}\{ 1,{\rm
exp}[(E_i-E_{i+1})(\beta_i-\beta_{i+1})]\}$, where $E_i$ is the energy
of the configuration initially at inverse temperature $\beta_i$.
For each sample we run $T_{\rm run}$ Metropolis sweeps on each 
configuration and then repeat the same procedure for $N_{\rm sample}$
different bond values. Note that, since we need to compute the overlap
parameter, we simulated at the same time two different replicas of the 
system.

Thermalization represents the main source of bias in simulations
of disordered systems. In all cases we start from 
a random infinite-temperature spin configuration.
To check for equilibration, we used the following method.
Let $\chi_i$ 
be the estimate of the overlap susceptibility $\chi_i$ at iteration $i$
at the largest $\beta$ value, $\beta_{\rm max}$, 
averaged over the $N_{\rm sample}$ disorder realizations. 
Then, define a block length $T_{\rm block}$ and the block-averaged
quantities
\begin{equation}
\chi_{b,i} = {1\over T_{\rm block}} 
\sum_{j=1}^{T_{\rm block}} \chi_{j + (i-1) T_{\rm block}}
\label{chibi}
\end{equation}
In our runs we typically choose $T_{\rm block} = 5000$ or 10000 
(all results presented below are always in units of Metropolis sweeps). 
In the tests we report below, in order to determine the thermalization
times more precisely, we use $T_{\rm block} = 2000$.
Finally, plot $\chi_{b,i}$ as a function of $i$. 
For $i$ large, $\chi_{b,i}$ becomes constant within error bars,
signalling equilibration. Data outside the approximately flat region are 
discarded in the calculation of the mean values. 

In the algorithm there are several parameters that must be tuned:
$\beta_{\rm min}$, $\beta_{\rm max}$, $N_\beta$, and $N_{\rm ex}$.
Moreover, $T_{\rm run}$ should be large enough to reach equilibration. 
In our simulations the difference $\Delta\beta$ between adjacent
$\beta$ values is kept constant, so that 
$\beta_{\rm max}-\beta_{\rm min}=(N_\beta-1)\Delta\beta$.  
The highest-temperature value $\beta_{\rm min}$ 
must be chosen such that, at this value of $\beta$,  the
standard Metropolis algorithm is reasonably efficient. 
In most of our simulations we use $\beta_{\rm min}=0.81$. 
At this value of $\beta$ (it corresponds to an infinite-volume 
correlation length 
$\xi_\infty \approx 3.7$) the Metropolis dynamics is reasonably fast.
Then, at fixed $\beta_{\rm min}$ and $\beta_{\rm max}$, we investigated
how the thermalization time depends on the two parameters
$N_\beta$ (or, equivalently, $\Delta\beta$, since we work
in a fixed $\beta$ interval) and $N_{\rm ex}$. 
The parameter $\Delta\beta$
controls the acceptance rate of the random-exchange moves and 
should not be too large, otherwise there are no temperature swaps
and the exchange dynamics becomes slow. 
As we shall see, $\Delta \beta$ is not a parameter that 
requires detailed tuning, since
there is a somewhat large 
interval of $\Delta\beta$ values for which the algorithm
is equally efficient, i.e. that all correspond to the 
minimal thermalization time (even though the acceptance 
rates change significantly). 
In order to perform a quantitative comparison,
we define the thermalization time  $T_{{\rm th},x\%}$
as the time such that $|\chi_{b,i}/\chi_{\rm ave} - 1| = x/100$,
where $i = T_{{\rm th},x\%}/T_{\rm block}$ and $\chi_{\rm ave}$
is the average value of $\chi$ at $\beta_{\rm max}$.
Since the typical errors on $\chi$ are $\approx 1\%$ we consider $x\%=2\%$.
The thermalization time  $T_{{\rm th},x\%}$ is of course
a lower bound on the true thermalization time, but it 
has the advantage of being computationally well-defined 
and thus it allows quantitative comparisons of the results 
corresponding to different choices of the parameters.
Moreover, we define a first round-trip time
$T_{\rm frt}$ (in units of Metropolis sweeps): it corresponds
to the time needed by a given configuration to go 
at least once through all temperatures.
For each simulation we collected
$2 N_\beta N_{\rm sample}$ first round-trip times.  Their distribution
is not Gaussian, but it has an exponential tail $e^{-c y}$. 
Therefore, instead of the average, we found more informative 
to consider the
time $T_{{\rm frt},90\%}$ such that 90\% of the 
measured $2N_\beta N_{\rm sample}$ first
round-trip times is smaller than $T_{{\rm frt},90\%}$.  
In our tests we found $T_{{\rm frt},90\%}$ to be 
related to the thermalization time. 
Equilibration is reached after a few $T_{{\rm frt},90\%}$.

\begin{figure*}[tb]
\centerline{\psfig{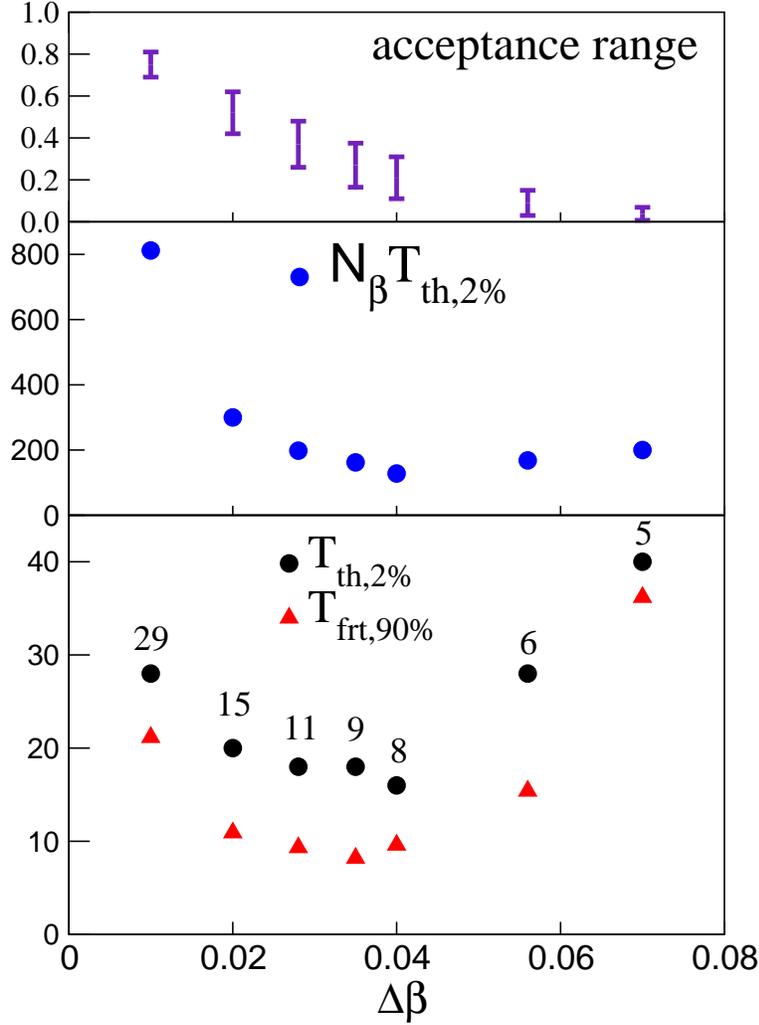}}
\vspace{2mm}
\caption{
We show the acceptance range (above), 
the quantity $N_\beta T_{\rm th,2\%}$ (in units of 
$1000$ sweeps) 
which is roughly proportional to the
computer time (middle), the 
thermalization time $T_{\rm th,2\%}$ and the first round-trip time 
$T_{\rm frt,90\%}$, both in units of
$1000$ sweeps, (below) versus $\Delta\beta$ for random-exchange
runs with $L=16$,
$\beta_{\rm min}=0.81$, $\beta_{\rm max}=1.09$, $N_{\rm ex}=10$.
The numbers reported close to the estimates of $T_{\rm th,2\%}$ correspond to
the number $N_\beta$ of $\beta$ values.
}
\label{raexl16}
\end{figure*}

In Fig.~\ref{raexl16}, for several values of $\Delta
\beta$, we show some data corresponding to 
simulations for $L=16$, $N_{\rm sample} = 1000$,
$\beta_{\rm min}=0.81$, $\beta_{\rm max}=1.09$,
and $N_{\rm ex}=10$. 
At the top we show the acceptance range. 
Since $\Delta \beta$ is kept fixed, the
acceptance rate is not constant. 
The minimum (maximum) value corresponds to swaps of the configurations with the 
smallest (largest) $\beta$.
The acceptance rate for the intermediate $\beta$
values varies approximately linearly.  In the figure we also
show the first round-trip time $T_{\rm frt,90\%}$, the thermalization time
$T_{\rm th,2\%}$ (in all cases we used $\chi_{\rm ave}=644(3)$ as estimate 
of $\chi$ at $\beta_{\rm max}=1.09$),
$T_{\rm comp}\approx N_\beta T_{\rm th,2\%}$, which is roughly
proportional to the computer time. 
The uncertainty on these quantities should be approximately 10\%.
The data show a significant correlation between the
thermalization and round-trip times. Their ratio is approximately
1.5-2, indicating that a few round trips are 
needed (and sufficient) to thermalize the system.
The thermalization time has a minimum in a relatively large region of
$\Delta \beta$ values, i.e., for $0.02 \lesssim \Delta \beta \lesssim 0.05$, 
corresponding to average acceptance rates 10-50\%.  This suggests that 
the thermalization time
$T_{\rm th}$ does not depend much on the acceptance rate as long 
as it is not too small (say, larger than 10\%).
If we consider the computer
time---and this is what we are really interested in---the optimal region 
is shifted to larger values of $\Delta \beta$. The minimum corresponds to 
approximately $\Delta \beta=0.04$, where the acceptance rate varies in
the interval $[0.115,0.312]$. However, an additional increase in 
$\Delta \beta$ does not worsen much the efficiency of the algorithm
that is nearly optimal even for $\Delta \beta=0.07$, where the 
acceptance rate is rather small (it varies between 0.7\% and 7\%).

The second interesting parameter is $N_{\rm ex}$.
We compare here
two simulations with $L=16$, $N_{\rm sample} = 1000$ 
using $N_{\rm ex}=10,20$ and keeping fixed the other parameters at
$\beta_{\rm min}=0.81$, $\beta_{\rm max}=1.09$, and $\Delta \beta=0.04$. 
We find 
$T_{\rm th,2\%}\approx 16000$ and 
$T_{\rm frt,90\%}\approx 9000$ for $N_{\rm ex}=10$,
and $T_{\rm th,2\%}\approx 28000$ and 
$T_{\rm frt,90\%}\approx 18000$ for $N_{\rm ex}=20$. 
Clearly, $N_{\rm ex} = 10$ is better than $N_{\rm ex}=20$.
This indicates 
that $N_{\rm ex}$ should be taken relatively small.  In our 
simulations we fixed $N_{\rm ex}=10$.

As expected, the thermalization time depends very strongly on 
$\beta_{\rm max}$.  For example, for
$L=16$, $\beta_{\rm min}=0.81$, and $\Delta \beta=0.01$, we find
$T_{\rm th,2\%}\approx 16000$ for $\beta_{\rm max}=1.00$ and
$T_{\rm th,2\%}\approx 28000$ for $\beta_{\rm max}=1.09$. Note that 
that the acceptance rates are similar varying 
from 0.686 (for $\beta = 0.81,0.82$), to
 0.772 (for $\beta = 0.99,1.00$) and
 0.805 (for $\beta = 1.08,1.09$).
We have performed an analogous test for 
$L=24$. For $\beta_{\rm min}=0.81$, 
$\Delta\beta=0.01$, $N_{\rm ex}=10$, 
we find 
$T_{{\rm frt},90\%}\approx 18000$ and 
$T_{\rm th,2\%}\lesssim 70000$ for $\beta_{\rm max}=1.0$,
$T_{{\rm frt},90\%}\approx 40000$ and
$T_{\rm th,2\%} > 100000$ for $\beta_{\rm max}=1.05$.

Finally, it is interesting to compare the thermalization
times for a Metropolis simulation without random-exchange moves 
and a random-exchange simulation.
For $L=16$ and $\beta = 1.0$ we performed a long 
Metropolis simulation, considering 5000 disorder realizations.
The thermalization time $T_{\rm th,2\%}$ is approximately
$400\cdot 10^3$. This should be compared with a random-exchange
run with $\beta \in [0.81,1.09]$, $\Delta\beta = 0.04$. In this 
case $T_{\rm th,2\%} \approx  16\cdot 10^3$. Even if the 
random-exchange run has a larger $\beta_{\rm max}$, the thermalization
time is much smaller.

\begin{table}[tbp]
\caption{ 
Our runs using the random-exchange method. Here $T_{\rm run}$
is the number of Metropolis sweeps per sample and $T_{\rm disc}$
is the number of discarded Metropolis sweeps.
In all cases $N_{\rm ex} = 10$.
}
\label{tablerun}
\begin{center}
\begin{tabular}{cclcrr}
\hline\hline
\multicolumn{1}{c}{$L$}&
\multicolumn{1}{c}{$\beta_{\rm min},\beta_{\rm max}$}&
\multicolumn{1}{c}{$\Delta\beta$}&
\multicolumn{1}{c}{$N_{\rm sample}$} &
\multicolumn{1}{c}{$T_{\rm run}$} &
\multicolumn{1}{c}{$T_{\rm disc}$} \\
\hline 
12 & 0.90,1.10 & 0.02 & 5000 & 50000 & 40000 \\
12 & 0.91,1.11 & 0.02 & 5000 & 50000 & 40000 \\
16 & 0.81,1.00 & 0.01 & 5000 & 50000 & 40000 \\
18 & 0.80,1.04 & 0.02 & 2000 & 100000 & 40000 \\
18 & 0.81,1.05 & 0.02 & 2000 & 100000 & 60000 \\
20 & 0.76,0.80 & 0.01 & 2000 & 50000 & 10000 \\
20 & 0.81,0.95 & 0.01 & 2000 & 100000 & 40000 \\
24 & 0.81,1.00 & 0.01 & 2000 & 100000 & 80000 \\
30 & 0.76,0.80 & 0.01 & 2000& 50000 & 16000 \\
30 & 0.81,0.95 & 0.01 & 1500& 100000 & 90000 \\
30 & 0.81,0.95 & 0.01 & 500 & 150000 & 90000 \\
\hline
\end{tabular}
\end{center}
\end{table}

\begin{figure*}[tb]
\centerline{\psfig{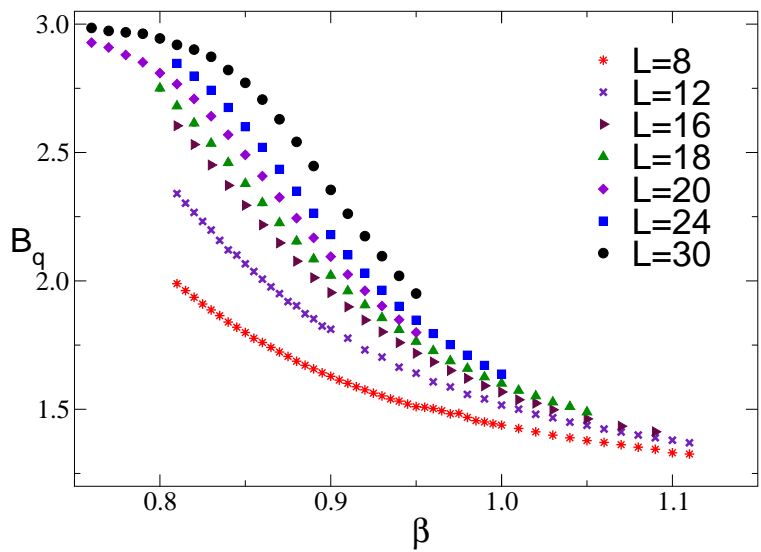}}
\vspace{2mm}
\caption{The quartic cumulant $B_q$ of the overlap parameter
for several lattice sizes.
}
\label{biq}
\end{figure*}

\begin{figure*}[tb]
\centerline{\psfig{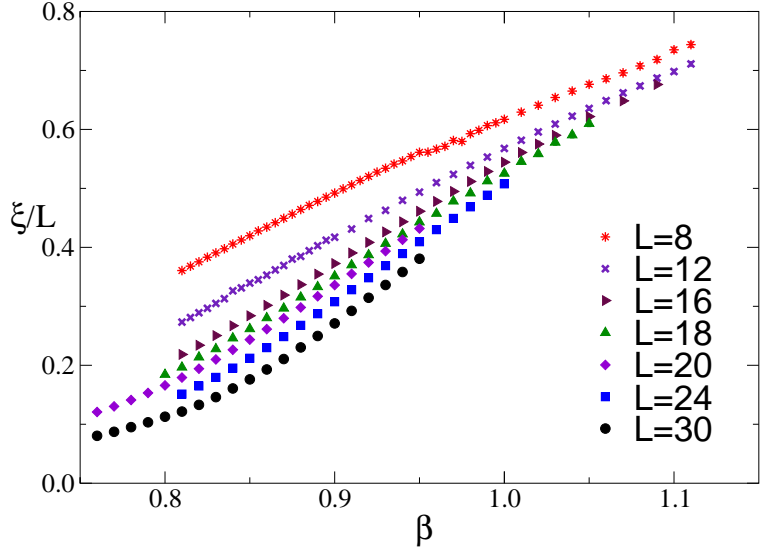}}
\vspace{2mm}
\caption{The ratio $\xi/L$ for several lattice sizes.
}
\label{xil}
\end{figure*}

\begin{figure*}[tb]
\centerline{\psfig{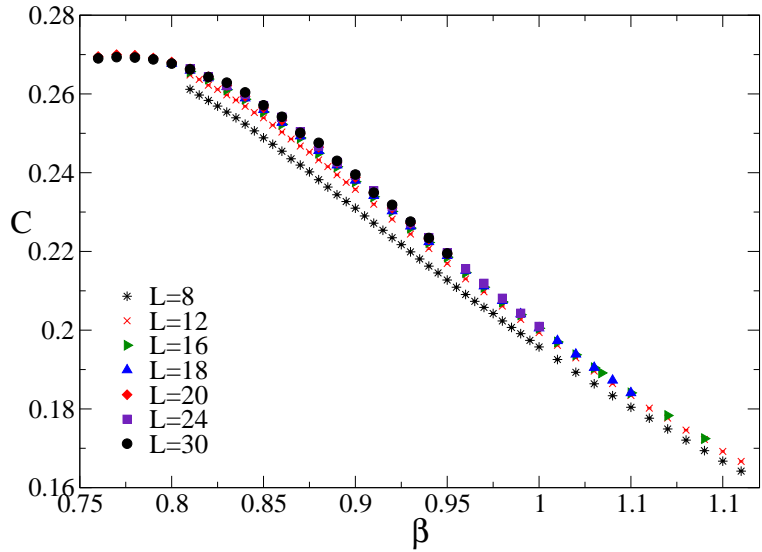}}
\vspace{2mm}
\caption{The specific heat for several lattice sizes.
}
\label{cv}
\end{figure*}

\begin{figure*}[tb]
\centerline{\psfig{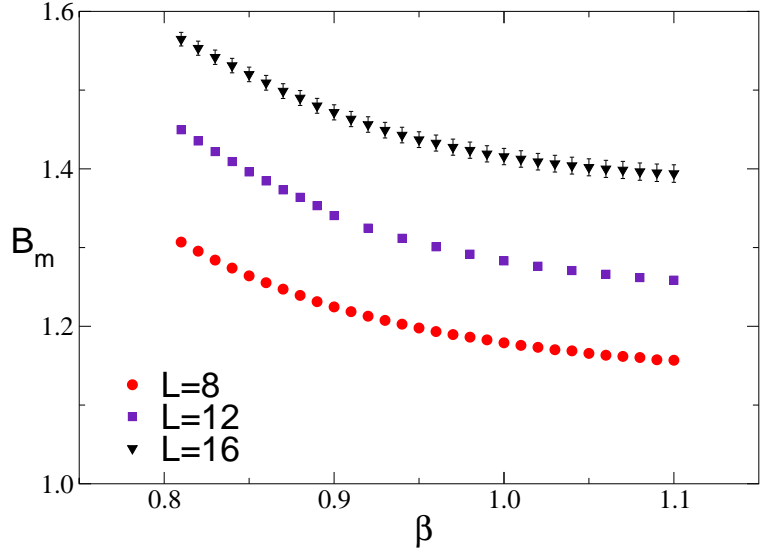}}
\vspace{2mm}
\caption{The magnetic quartic cumulant $B_m$ for several lattice sizes.
}
\label{Bim}
\end{figure*}

\section{Results}
\label{res}

We consider model (\ref{raminfD}) on simple cubic lattices $L^3$ with periodic
boundary conditions.  We have performed a series of simulations for lattices
with $8 \le L \le 30$, using the random-exchange method, as explained in
Sec.~\ref{mcsim}. For each value of $L$ we considered a range of $\beta$
values such that the thermalization time was less than $10^5$ Metropolis
sweeps per sample. We could therefore take $T_{\rm run}\approx 10^5$. We
required this condition in order to be able to have $N_{\rm sample} \ge 1000$,
and thus precise estimates. Of course, this limits the parameter $\beta_{\rm
  max}$ and, for $L=30$, we were able to collect data only up to $\beta =
0.95$.  The parameters of our random-exchange runs are reported in
Table~\ref{tablerun}.  We are not reporting several other runs with $L=16$,
$N_{\rm sample} = 1000$, that have been used in the numerical tests reported
in Sec.~\ref{mcsim}. Few runs with $L=8,12,16$ were performed
by using only Metropolis updatings without random-exchange moves. In
Table~\ref{tablerun} we report the length $T_{\rm run}$ of the run for each
value of the parameters and $T_{\rm disc}$, the number of sweeps that have
been discarded before measuring (this parameter has been chosen conservatively
to avoid any thermalization bias).  Simulations took approximately 2.5 CPU
years of a workstation equipped with an AMD Opteron Processor 246 (2 GHz
clock).

In Figs.~\ref{biq} and \ref{xil} we show the quartic cumulant $B_q$
defined in Eq.~(\ref{binder}) and the ratio $\xi/L$ 
for several lattice sizes $L$.  In the
region of $L$ and $\beta$ covered by our MC data, neither $B_q$ nor 
$\xi/L$ have a crossing point. Thus, we do not have direct evidence 
for a finite-temperature transition in the range of temperatures that we 
can reliably simulate.
No indication of a phase transition in this range of $\beta$ is 
also provided 
by the results for the specific heat---they are shown in 
Fig.~\ref{cv}---and by magnetic variables---there is no indication of 
a crossing point in the results for $B_m$ defined in Eq.~(\ref{binders}),
see Fig.~\ref{Bim}.

These results are consistent with two possible scenarios: (i) the system
becomes critical at $\beta = \beta_c$ with $\beta_c \gtrsim 1.05$ (with the 
possibility of a zero-temperature transition, i.e. 
$\beta_c = +\infty)$; (ii) the system never shows criticality
and even for $\beta = \infty$ the correlation length is finite. We will now
show that our data allow us to exclude this second possibility since,
as $\beta$ increases, the model shows critical behavior; more precisely,
our data for the overlap variables 
show FSS as expected close to a critical point. In order to
make the check as reliable as possible we will study the FSS behavior of
the ratios $A(\beta,s L)/A(\beta,L)$, where $A(\beta,L)$ is a long-distance 
quantity and $s$ is an arbitrary number. In the FSS limit 
(i.e. for $L,\xi(\beta,L)\rightarrow \infty$ at $\xi(\beta,L)/L$ fixed) 
we should have
\begin{equation}
{A(\beta,sL)\over A(\beta,L)} \longrightarrow
   F_A\left[s,\xi(\beta,L)/L\right] ,
\label{stl3}
\end{equation}
where $F_A(s,z)$ is a universal function. Note that in this formulation
there are no free parameters and thus one can make an unbiased test of FSS.

The FSS curves computed by using Eq.~(\ref{stl3})
can be used to determine infinite-volume
quantities for large values of the correlation length. For this purpose
we employ the extrapolation
method of \cite{CEFPS-95} (see also \cite{CEPS-95,MPS-97} for
a discussion of the efficiency of this technique). Indeed, in the absence of
scaling corrections, Eq.~(\ref{stl3}) 
allows us to compute $A(\beta,sL)$ on a
lattice of size $sL$ in terms of quantities defined on a lattice of
size $L$ and of the function $F_A(s,z)$. In practice, one works as
follows.  First, one performs several runs, determining $A(\beta,sL)$,
$A(\beta,L)$, $\xi(\beta,sL)$, and $\xi(\beta,L)$.  By means of a
suitable interpolation, this provides the FSS function $F_A(s,z)$ for $A$ and
$\xi$.  Then, $A_\infty(\beta)$ and $\xi_\infty(\beta)$ are obtained
from $A(\beta,L)$ and $\xi(\beta,L)$ by iterating Eq.~(\ref{stl3}) and
the corresponding equation for $\xi(\beta,L)$.  Of course, one must be
very careful about scaling corrections, discarding systematically
lattices with small values of $L$ till results become independent
of $L$ within error bars.

\begin{figure*}[tb]
\centerline{\psfig{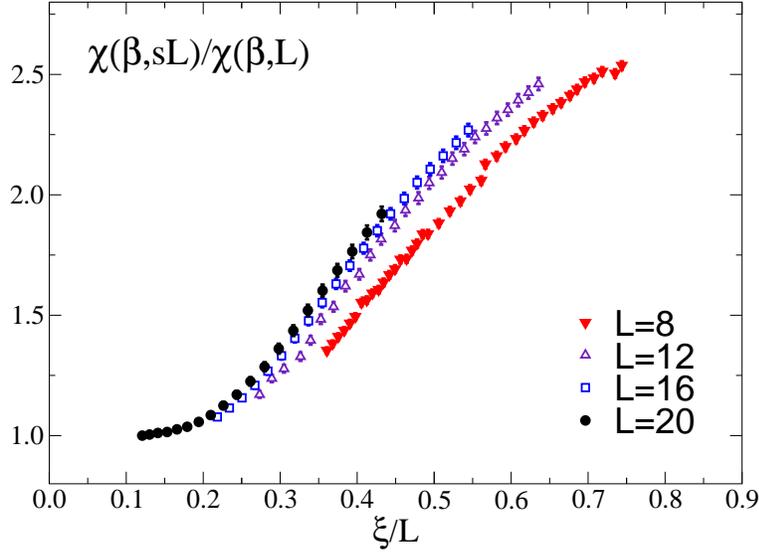}}
\vspace{2mm}
\caption{
The FSS curve of the susceptibility $\chi$ for $s=3/2$.
}
\label{chi}
\end{figure*}

\begin{figure*}[tb]
\centerline{\psfig{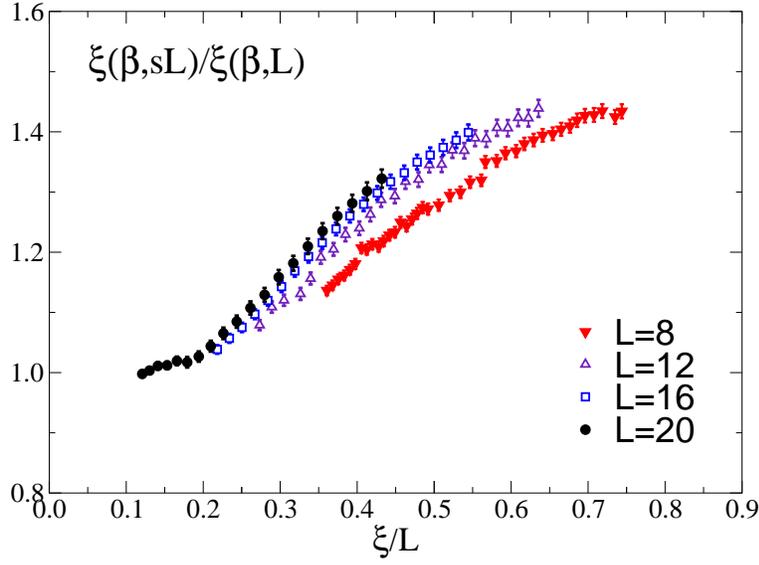}}
\vspace{2mm}
\caption{
The FSS curve of the correlation length $\xi$ for $s=3/2$. 
}
\label{xi}
\end{figure*}

\begin{figure*}[tb]
\centerline{\psfig{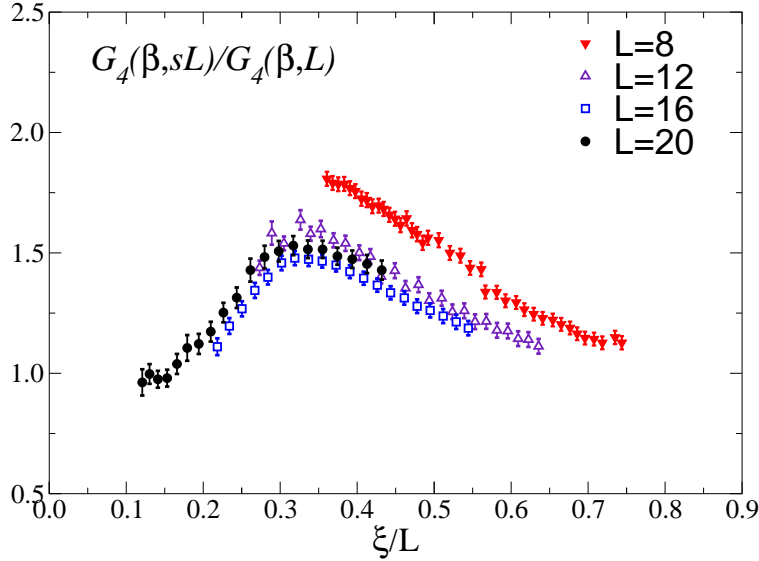}}
\vspace{2mm}
\caption{
The FSS curve of the zero-momentum quartic coupling $G_4$ for $s=3/2$.
}
\label{G4}
\end{figure*}

\begin{figure*}[tb]
\centerline{\psfig{width=10truecm,angle=0,file=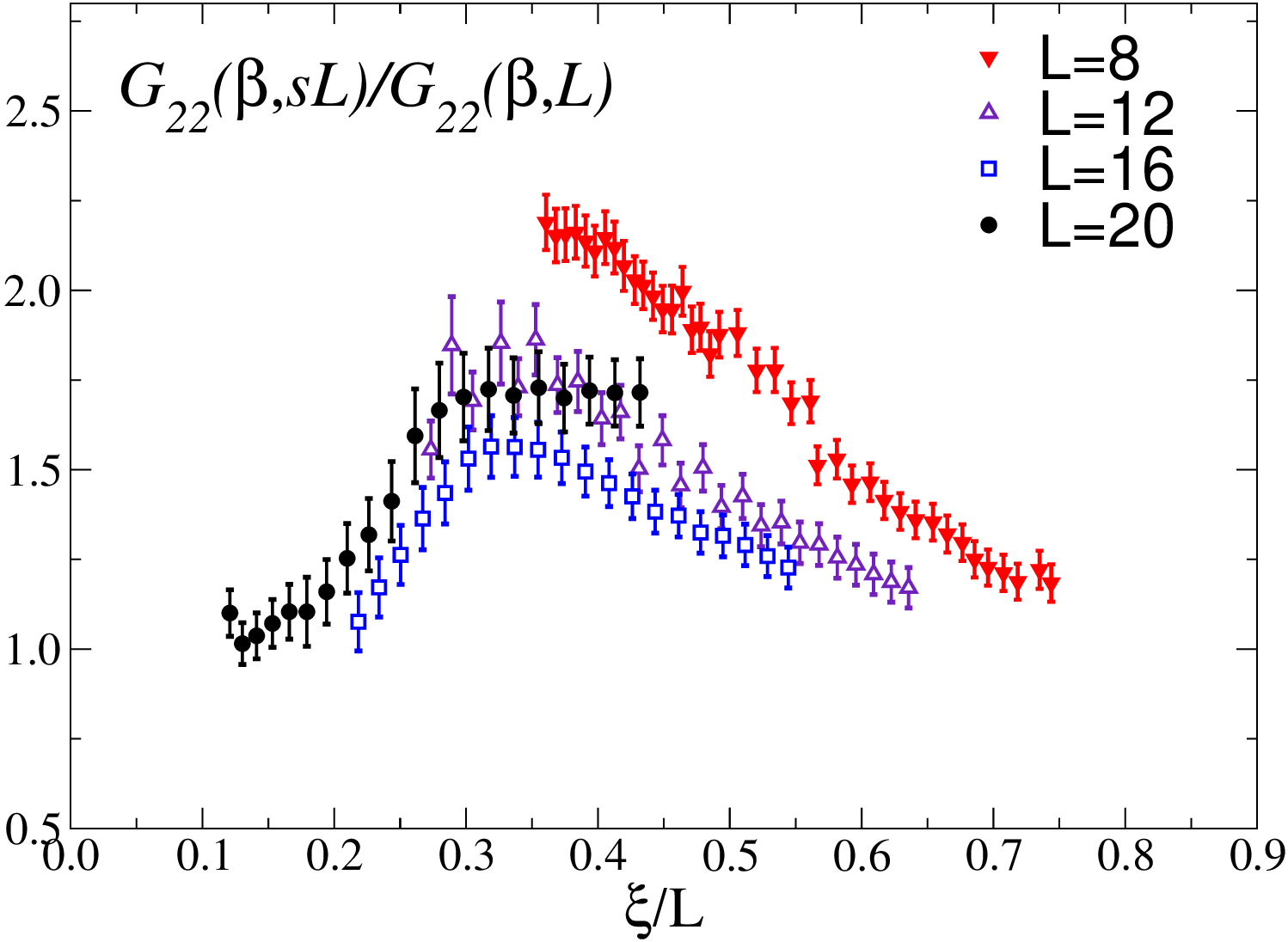}}
\vspace{2mm}
\caption{
The FSS curve of the zero-momentum quartic coupling $G_{22}$ for $s=3/2$.
}
\label{G22}
\end{figure*}

In Figs.~\ref{chi}, \ref{xi}, \ref{G4}, and \ref{G22} we plot
the ratios (\ref{stl3}) for $\chi$, $\xi$, $G_4$, and $G_{22}$, 
fixing $s = 3/2$. The curves for $\chi$ and $\xi$ show significant scaling
corrections. However, they apparently decrease very rapidly
with $L$ and indeed the data corresponding to the pairs $L=16,24$ are 
only slightly different from those with $L=20,30$.
Therefore, they strongly suggest that FSS holds, although,
for $L\lesssim 30$, scaling corrections are significant
compared to our error bars (for $\chi$ our data have a relative
error of approximately 1\% for  $L=20$ and $L=30$). For $G_4$ and $G_{22}$
corrections are apparently weaker and indeed the data corresponding
to $L=20,30$ are compatible with those with $L=12,18$ and with $L=16,24$.
In this case FSS holds within error bars.

The presence of scaling corrections in the FSS curve for $\xi$ 
does not allow us to use straightforwardly the iteration method of 
\cite{CEFPS-95} and makes it necessary to include scaling 
corrections in the scaling Ansatz. As done in \cite{CMPV-03}, 
we use a more general Ansatz of the form
\begin{equation}
{A(\beta,sL)\over A(\beta,L)} = 
  F_A(s,\xi(\beta,L)/L) + L^{-\omega} G_A(s,\xi(\beta,L)/L),
\label{FSS-corr}
\end{equation}
where $G_A(s,z)$ is a new universal scaling function and 
$\omega$ is a to-be-determined correction-to-scaling exponent.

Our data are not precise enough to allow a determination of $\omega$.
However, it is easy to convince oneself that $\omega$ cannot 
be arbitrarily large. If we consider the correlation length
$\xi(\beta,L)$, the correction-to-scaling exponent should be 
less than 2, since corrections with exponent 2 appear 
necessarily \cite{CP-98}. They are related to the very definition of the 
second-moment correlation length we use on a finite lattice. 
Indeed, it corresponds to its infinite-volume counterpart only
up to terms of order $L^{-2}$. For $\chi$, $G_{4}$, and $G_{22}$
there are corrections due to the analytic background 
\cite{Privman,SS-99}
that are proportional to $L^{2-\eta_o}$, where $\eta_o$ is the 
susceptibility exponent, $\chi \sim \xi^{2-\eta_o}$. 
As we shall see, 
$\eta_o \approx -0.3$, so that $\omega \lesssim 2.3$. 
On the other hand, one cannot set {\em a priori} a lower bound 
on $\omega$ and, in principle, $\omega$ can be arbitrarily small.
We will present below analyses with $\omega$ as small as $0.2$,
presenting results for 
$\omega = 0.2$, $\omega = 1$, and $\omega = 2$. 
This choice will be justified below. 

\begin{figure*}[tb]
\centerline{\psfig{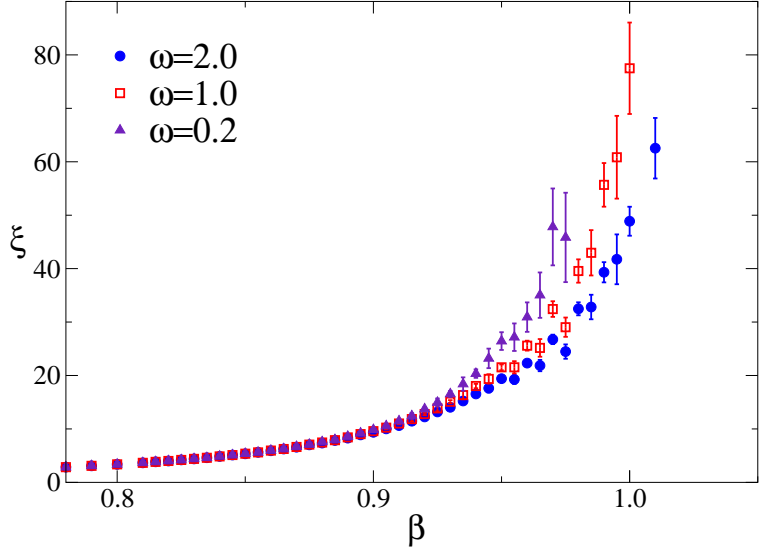}}
\vspace{2mm}
\caption{
The infinite-volume correlation length $\xi_\infty$ as obtained by
the extrapolation FSS method.
}
\label{xivsbeta}
\end{figure*}

\begin{figure*}[tb]
\centerline{\psfig{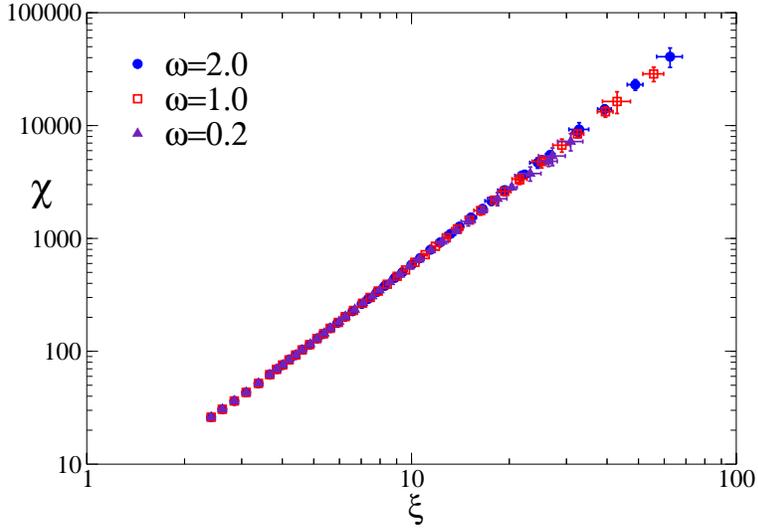}}
\vspace{2mm}
\caption{
Log-log plot of the infinite-volume estimates of $\chi_\infty$ versus 
$\xi_\infty$.
}
\label{chivsxi}
\end{figure*}

In order to determine $\eta_o$, we consider the data reported in 
Figs.~\ref{chi} and \ref{xi} and apply the extrapolation technique
using Eq.~(\ref{FSS-corr}).
As in \cite{CEFPS-95} the scaling curves are parametrized
as polynomials in $\exp[-\xi(\beta,L)/L]$ satisfying 
$G_A(s,0) = F_A(s,0) = 1$. A fourth-order polynomial for $F_A(s,z)$
and a third-order polynomial for $G_A(s,z)$ appear to be adequate.
In Fig.~\ref{xivsbeta} we show the infinite-volume correlation
length $\xi_\infty(\beta)$. It can be determined 
reliably up to $\beta = 0.95$, essentially because we have 
results for the largest lattice, $L = 30$, only up to this value of 
$\beta$. The dependence on $\omega$ is not large, but in any case larger
that the statistical errors. For $\beta = 0.95$, we find 
$\xi_\infty = 26.4(1.7)$, $21.5(5)$, $19.4(4)$ for 
$\omega = 0.2$, 1, 2, respectively. Whatever is the value of 
$\omega$, $\xi_\infty(\beta)$ increases quite rapidly,
confirming that the system eventually becomes critical. 
Evidence of criticality is also provided in Fig.~\ref{chivsxi} 
where we show a log-log plot of $\chi_\infty$ versus $\xi_\infty$. 
The behavior is linear, indicating that 
\begin{equation}
\chi_\infty \sim \xi_\infty^{2-\eta_o},
\label{chioxi}
\end{equation}
where $\eta_o$ is a critical exponent. Note that the dependence
on $\omega$ of $\chi_\infty$ at fixed $\xi_\infty$ is much smaller than 
the $\omega$ dependence 
of $\chi_\infty$ and $\xi_\infty$ at fixed $\beta$. This is due to the fact 
that, at fixed $\beta$, $\chi_\infty$ and $\xi_\infty$ are strongly
correlated: they both increase with 
decreasing $\omega$, in such a way that the ratio 
$\chi_\infty/\xi_\infty^{2-\eta_o}$
has a tiny dependence on $\omega$.  To obtain an estimate of 
$\eta_o$ we fitted $\chi_\infty$ and $\xi_\infty$ to $\ln \chi_\infty =
a + (2 - \eta_o) \ln \xi_\infty$. To estimate the scaling corrections,
the fit has been repeated several times, each time including only the data 
satisfying $\beta\ge \beta_{\rm min}$. 
Results are reported in Table~\ref{table:etaest} for different values 
of $\omega$. In order to have an additional check on the 
stability of the results, we have also repeated the analysis twice:
we present results obtained by  using 
all data ($L_{\rm min} = 8$) and results 
obtained using only lattices with
$L\ge 12$ ($L_{\rm min} = 12$).

\begin{table}
\caption{Estimates of the exponent $\eta_o$ for several 
values of $\omega$ and $L_{\rm min}$ .}
\label{table:etaest}
\begin{center}
\begin{tabular}{cllllll}
\hline\hline
  & \multicolumn{3}{c}{$L_{\rm min} = 8$} 
  & \multicolumn{3}{c}{$L_{\rm min} = 12$} \\
$\beta_{\rm min}$ & \multicolumn{1}{c}{$\omega=0.2$} & 
            \multicolumn{1}{c}{$\omega=1$} & 
            \multicolumn{1}{c}{$\omega=2$} & 
            \multicolumn{1}{c}{$\omega=0.2$} & 
            \multicolumn{1}{c}{$\omega=1$} & 
            \multicolumn{1}{c}{$\omega=2$} \\ \hline
 0.76 & $-$0.161(7)&  $-$0.177(6)&  $-$0.185(6)&
    $-$0.148(9)&  $-$0.159(8)&  $-$0.167(7)\\
 0.78 & $-$0.191(9)&  $-$0.203(8)&  $-$0.214(7)&
    $-$0.175(11)&  $-$0.185(9)&  $-$0.196(9)\\
 0.80 & $-$0.210(12)&  $-$0.221(10)&  $-$0.233(9)&
    $-$0.196(15)&  $-$0.203(12)&  $-$0.216(11)\\
 0.82 & $-$0.221(15)&  $-$0.234(12)&  $-$0.247(10)&
    $-$0.205(21)&  $-$0.214(15)&  $-$0.227(14)\\
 0.84 & $-$0.236(22)&  $-$0.251(16)&  $-$0.268(14)&
    $-$0.225(31)&  $-$0.234(22)&  $-$0.242(19)\\
 0.86 & $-$0.233(32)&  $-$0.258(22)&  $-$0.276(19)&
    $-$0.217(50)&  $-$0.233(31)&  $-$0.251(27)\\
 0.88 & $-$0.236(50)&  $-$0.265(31)&  $-$0.288(26)&
    $-$0.213(81)&  $-$0.241(45)&  $-$0.262(38)\\
\hline
\end{tabular}
\end{center}
\end{table}

The results presented in Table~\ref{table:etaest} depend 
somewhat on $\omega$ and $\beta_{\rm min}$. At fixed 
$\omega$, $\eta_o$ decreases with increasing $\beta_{\rm min}$,
reaching an approximate plateau within error bars 
for $\beta_{\rm min}\gtrsim 0.84$. At fixed $\beta_{\rm min}$
the estimates decrease with increasing $\omega$. A conservative estimate
can be obtained by noting that all results with 
$0.84 \lesssim \beta_{\rm min}\lesssim 0.86$---for larger
values of $ \beta_{\rm min}$ error bars are quite large---lie
in the interval $-0.28\lesssim \eta_o \lesssim -0.18$, 
including statistical errors. We thus end up with 
the following estimate:
\begin{equation}
\eta_o = - 0.23(5).
\end{equation}
\begin{figure*}[tb]
\centerline{\psfig{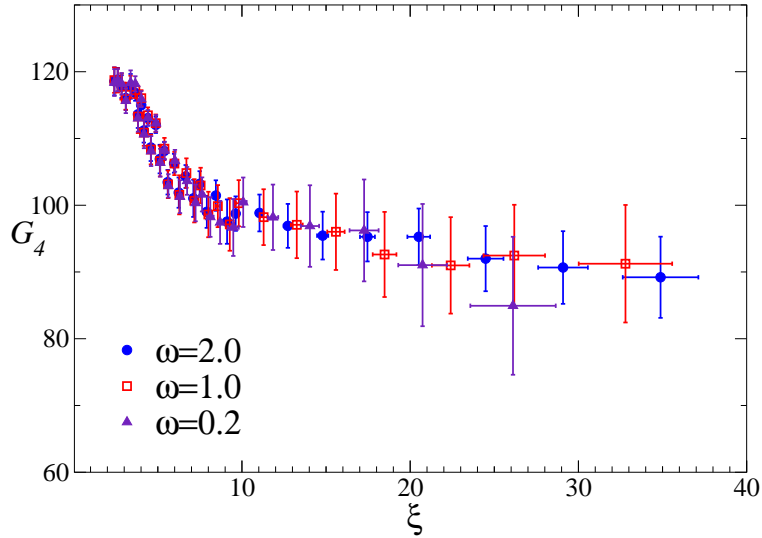}}
\vspace{2mm}
\caption{
Estimates of $G_{4,\infty}$ versus $\xi_\infty$ for three different
values of $\omega$.
}
\label{G4vsxi}
\end{figure*}
\begin{figure*}[tb]
\centerline{\psfig{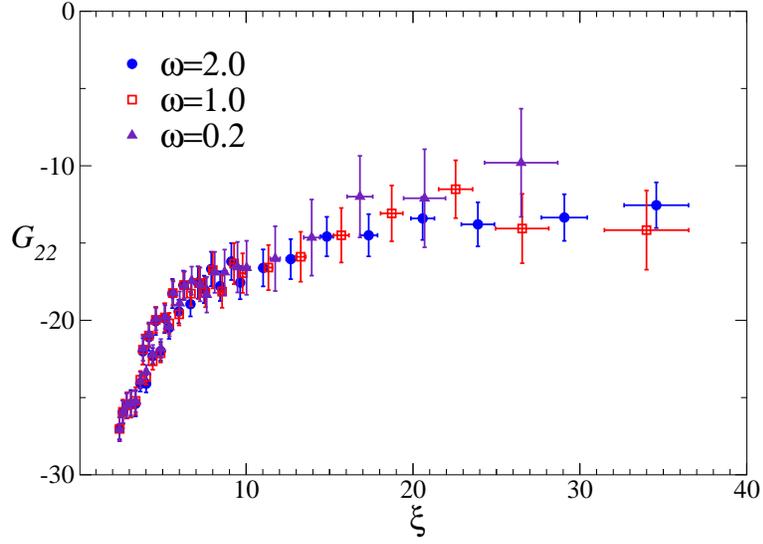}}
\vspace{2mm}
\caption{
Estimates of $G_{22,\infty}$ versus $\xi_\infty$ for three different
values of $\omega$.
}
\label{G22vsxi}
\end{figure*}
Then, we consider the RG-invariant quantities $G_4$ and $G_{22}$.
Again, we use Eq.~(\ref{FSS-corr}) and 
the iteration algorithm to determine infinite-volume 
estimates $G_{4,\infty}(\beta)$ and 
$G_{22,\infty}(\beta)$.
Reasonable results are only obtained if the data
with $L = 8$ are discarded, i.e. if we only consider 
$L\ge L_{\rm min} = 12$. Indeed, if all MC data are 
used in the FSS extrapolation,  $G_{4,\infty}(\beta)$ and
$G_{22,\infty}(\beta)$ apparently do not converge to 
a finite value as $\beta\to\infty$.
For $L_{\rm min} = 12$, the $\omega$ dependence is of the 
order of the statistical error bars: for instance,
for $\beta= 0.95$ we predict 
$G_{4,\infty} = 85(10)$, 91(7), 95(4), and
$G_{22,\infty} = -13.4(7)$, $-13.1(7)$, $-11(2)$,
for $\omega = 0.2$, 1, 2. 
The results are shown in
Figs.~\ref{G4vsxi} and \ref{G22vsxi}.
They show a rapid variation with $\xi_\infty$ for 
small values of the correlation length and reach a
plateau for $\xi_\infty\gtrsim 15$. In order to estimate
the critical value for $\xi_\infty\to\infty$, we perform 
a fit of the form
\begin{equation}
{\cal R}_{\infty} = {\cal R}^* + {a_{\cal R}\over \xi_\infty^\Omega},
\label{Rfit}
\end{equation}
where ${\cal R}^*$, $a_{\cal R}$, and $\Omega$ are free parameters,
${\cal R} = G_4$ or $G_{22}$. Results are reported in 
Tables \ref{table:G4est} and \ref{table:G22est}. 
The dependence on $\omega$ is here quite small, as it has 
to be expected since there is no evidence of scaling 
corrections in the FSS curves for $G_4$ and $G_{22}$. 
The dependence we observe here is mainly related to the 
corrections affecting the correlation length that is also 
used in the extrapolation. Again our final estimate is 
obtained conservatively by looking at the variation 
with $\omega$ and $\beta_{\rm min}$. For $G_4^*$, we take as 
our final estimate 
\begin{equation}
G_4^* = 88(8),
\label{G4est}
\end{equation}
where the error is such to include all results with 
$\beta_{\rm min}\ge 0.81$. Analogously, we estimate
\begin{equation}
G_{22}^* = -11(4).
\end{equation}
The fit parameter $\Omega$ that appears in Eq.~(\ref{Rfit})
is not completely arbitrary and can be related to the 
exponent $\omega$ used in the FSS extrapolation.
Thus, the results of 
Tables \ref{table:G4est} and \ref{table:G22est} can be used
to obtain constraints on the value of $\omega$. In general, given 
an infinite-volume  RG-invariant quantity ${\cal R}(\beta)$,
close to a critical point we expect an expansion of the form:
\begin{equation}
{\cal R}(\beta) = {\cal R}^*(1  + a |\beta - \beta_c|^\Delta),
\label{Rcorr}
\end{equation}
with a positive correction-to-scaling exponent $\Delta$.
According to the RG theory,
scaling corrections may have several origins 
\cite{Wegner-76,AF-82}:
\begin{itemize}
\item[(i)] There are analytic corrections  of the form
$|\beta - \beta_c|^n$, $n$ being an integer. 
\item[(ii)] There are nonanalytic corrections related to the 
irrelevant operators. They have the 
form $|\beta - \beta_c|^{n \nu \omega_i}$,
$|\beta - \beta_c|^{n \nu \omega_i + m \nu \omega_j}$, etc., where
$\omega_i=-y_i$ and $y_i$ are the RG dimensions of the irrelevant
RG operators, and $n$, $m$ are integers.
\item[(iii)] There are corrections related to the analytic 
background. For instance, in the susceptibility there are 
corrections proportional to $|\beta - \beta_c|^\gamma$. 
\end{itemize}
In \cite{PV-98} an argument was given to show
that renormalized coupling constants, like $G_4$ and $G_{22}$,
 do not have analytic
corrections. The absence of this type of corrections was 
verified in the continuum $O(N)$ $\phi^4$ theory in $d$ dimensions
to order $1/N$ \cite{PV-98} and in the two-dimensional 
Ising model \cite{CCCPV-00,CHPV-02}. In the absence of analytic 
corrections, the leading contribution is due to the 
irrelevant operators, and therefore $\Delta = \omega_1 \nu$,
where $\omega_1 = - y_1$ and $y_1$
 is the RG dimension of the leading irrelevant
operator in the model. Therefore, in Eq.~(\ref{Rfit}) the exponent
$\Omega$ should be identified with $\omega_1$.

\begin{table}
\caption{Estimates of $G_4^*$ and $\Omega$ for different values of 
$\omega$. Here $L_{\rm min} = 12$.}
\label{table:G4est}
\begin{center}
\begin{tabular}{cllllll}
\hline\hline
  & \multicolumn{2}{c}{$\omega = 0.2$} 
  & \multicolumn{2}{c}{$\omega = 1$} 
  & \multicolumn{2}{c}{$\omega = 2$}  \\
$\beta_{\rm min}$ &
  \multicolumn{1}{c}{$G_4^*$} & \multicolumn{1}{c}{$\Omega$} &
  \multicolumn{1}{c}{$G_4^*$} & \multicolumn{1}{c}{$\Omega$} &
  \multicolumn{1}{c}{$G_4^*$} & \multicolumn{1}{c}{$\Omega$} \\
\hline
 0.780 &    58(26)&   0.36(19)& 
          72(13)&   0.47(17)& 
          78(8)&   0.55(15)\\ 
 0.790 &    74(13)&   0.58(23)& 
          78(9)&   0.58(19)& 
          83(6)&   0.70(18)\\ 
 0.800 &    85(7)&   0.95(30)& 
          86(6)&   0.89(25)& 
          87(4)&   0.93(21)\\ 
 0.810 &    87(6)&   1.08(34)& 
          87(5)&   1.01(29)& 
          88(4)&   0.98(23)\\ 
 0.815 &    86(7)&   1.00(38)& 
          88(5)&   1.07(34)& 
          88(4)&   0.98(27)\\ 
 0.820 &    89(6)&   1.25(45)& 
          89(5)&   1.20(38)& 
          89(4)&   1.07(29)\\ 
 0.825 &    88(7)&   1.13(50)& 
          88(6)&   1.06(40)& 
          88(5)&   1.01(32)\\ 
 0.830 &    91(6)&   1.47(60)& 
          90(5)&   1.34(49)& 
          90(4)&   1.20(38)\\ 
 0.835 &    91(6)&   1.49(76)& 
          89(6)&   1.18(56)& 
          89(5)&   1.07(43)\\ 
 0.840 &    95(4)&   2.49(1.25)& 
          92(5)&   1.66(78)& 
          91(4)&   1.37(54)\\ 
 0.845 &    89(10)&   1.13(95)& 
          86(11)&   0.82(65)& 
          84(11)&   0.70(51)\\ 
 0.850 &    91(8)&   1.45(1.20)& 
          87(9)&   0.94(73)& 
          85(10)&   0.76(55)\\ 
\hline
\end{tabular}
\end{center}
\end{table}

\begin{table}
\caption{Estimates of $G_{22}^*$ and $\Omega$ for different values of
$\omega$. Here $L_{\rm min} = 12$.}
\label{table:G22est}
\begin{center}
\begin{tabular}{cllllll}
\hline\hline
  & \multicolumn{2}{c}{$\omega = 0.2$} 
  & \multicolumn{2}{c}{$\omega = 1$} 
  & \multicolumn{2}{c}{$\omega = 2$}  \\
$\beta_{\rm min}$ &
  \multicolumn{1}{c}{$G_{22}^*$} & \multicolumn{1}{c}{$\Omega$} &
  \multicolumn{1}{c}{$G_{22}^*$} & \multicolumn{1}{c}{$\Omega$} &
  \multicolumn{1}{c}{$G_{22}^*$} & \multicolumn{1}{c}{$\Omega$} \\
\hline
 0.780 &   $-$8.1(3.8)&   0.69(22)& 
        $-$10.6(1.8)&   0.81(16)& 
        $-$10.1(1.5)&   0.76(13)\\ 
 0.790 &  $-$10.2(3.0)&   0.88(27)& 
        $-$11.5(1.6)&   0.94(19)& 
        $-$10.9(1.3)&   0.88(16)\\ 
 0.800 &  $-$11.1(2.8)&   1.00(32)& 
        $-$11.8(1.5)&   1.01(22)& 
        $-$11.3(1.3)&   0.94(18)\\ 
 0.810 &  $-$10.6(3.3)&   0.92(34)& 
        $-$11.7(1.6)&   0.99(24)& 
        $-$11.1(1.4)&   0.90(19)\\ 
 0.815 &  $-$10.1(3.8)&   0.86(37)& 
        $-$11.8(1.7)&   1.01(27)& 
        $-$10.9(1.6)&   0.87(21)\\ 
 0.820 &  $-$11.6(2.9)&   1.10(44)& 
        $-$12.3(1.5)&   1.14(29)& 
        $-$11.6(1.3)&   1.01(24)\\ 
 0.825 &   $-$9.1(5.4)&   0.73(44)& 
        $-$11.5(2.0)&   0.93(30)& 
        $-$10.2(2.1)&   0.75(24)\\ 
 0.830 &  $-$11.2(3.6)&   1.03(54)& 
        $-$12.4(1.6)&   1.18(37)& 
        $-$10.9(1.8)&   0.87(27)\\ 
 0.835 &   $-$9.6(5.6)&   0.78(55)& 
        $-$11.5(2.3)&   0.91(38)& 
        $-$10.0(2.7)&   0.71(30)\\ 
 0.840 &  $-$12.8(2.9)&   1.53(93)& 
        $-$12.6(1.7)&   1.25(49)& 
        $-$11.1(2.0)&   0.90(35)\\ 
 0.845 &   $-$9.5(7.2)&   0.73(71)& 
        $-$10.2(4.0)&   0.67(45)& 
         $-$7.9(5.7)&   0.49(36)\\ 
 0.850 &  $-$10.0(6.6)&   0.81(80)& 
        $-$10.3(4.0)&   0.68(47)& 
         $-$8.1(5.7)&   0.51(38)\\ 
\hline
\end{tabular}
\end{center}
\end{table}

Corrections appearing in Eq.~(\ref{Rcorr}) are strictly related 
to corrections appearing in FSS. In the FSS case no analytic 
corrections are expected, not only in $G_{4}$ or $G_{22}$
but in any quantity.\footnote{A careful discussion is presented in 
\cite{SS-99}.  The absence of analytic corrections in FSS
is shown in \cite{GJ-87}
and has been checked to order $L^{-2}$ in the
two-dimensional Ising model  \cite{CHPV-02}
and to order $L^{-1}$ in the three-dimensional
XY model \cite{CHPV-prep}.}
Corrections of type (ii) correspond to terms of the 
form $L^{-n \omega_i}$, $L^{-n \omega_i - m\omega_j}$, so that 
the leading correction in FSS has the form $L^{-\omega_1}$. Thus,
the exponent $\Omega$ should be identified with 
the exponent that controls FSS corrections and has been 
used in the FSS Ansatz (\ref{FSS-corr}). The results reported
in Tables~\ref{table:G4est} and \ref{table:G22est} 
give estimates of $\Omega$ that show a tiny dependence on $\omega$
and we can safely estimate
\begin{equation}
\Omega = 1.0(4).
\end{equation}
This result justifies {\em a posteriori} our choice of not considering 
values of $\omega$ smaller than 0.2. The estimate of $\Omega$ 
restricts the interval in which $\omega$ can vary. 
This allows us to obtain more precise estimates 
for $\eta_o$ and $G_{22}^*$ (the estimate of 
$G_4^*$ does not vary since this quantity has a very
small dependence on $\omega$).
By assuming $0.6\le \omega \le 1.4$ we obtain the results:
\begin{eqnarray}
&& \eta_o = -0.24(4), \\
&& G_{22}^* = -11.5(2.5).
\end{eqnarray}
\begin{table}[tbp]
\caption{ 
Estimates of $\beta_c$ and $\gamma$ (second and third column) obtained from
the fit of $\chi_\infty$ and of $\beta_c$ and $\nu$ (fourth and 
fifth column) obtained from the fit of $\xi_\infty$.
The infinite-volume estimates of $\chi_\infty$ and  $\xi_\infty$
have been obtained by applying 
the extrapolation method based on Eq.~(\ref{FSS-corr}) with 
$\omega = 1$ to all data ($L_{\rm min} = 8$).
}
\label{fit-omega1p0}
\begin{center}
\begin{tabular}{lllll}
\hline\hline
& \multicolumn{2}{c}{fit of $\chi_\infty$} 
& \multicolumn{2}{c}{fit of $\xi_\infty$} \\
\multicolumn{1}{c}{$\beta_{\rm min}$}&
\multicolumn{1}{c}{$\beta_c$}&
\multicolumn{1}{c}{$\gamma_o$}  &
\multicolumn{1}{c}{$\beta_c$} &
\multicolumn{1}{c}{$\nu_o$}   \\
\hline 
0.76 &  1.060(2)&   4.78(4)&  1.083(3)&   2.43(4)\\
 0.78 &  1.059(2)&   4.76(5)&  1.075(4)&   2.33(4)\\
 0.80 &  1.060(2)&   4.76(6)&  1.073(4)&   2.29(5)\\
 0.82 &  1.062(3)&   4.84(8)&  1.073(5)&   2.30(6)\\
 0.84 &  1.064(4)&   4.89(12)&  1.072(6)&   2.28(8)\\
 0.86 &  1.070(6)&   5.11(19)&  1.076(9)&   2.35(13)\\
 0.88 &  1.074(9)&   5.27(32)&  1.080(13)&   2.42(21)\\
 0.90 &  1.082(16)&  5.60(62)&  1.080(19)&   2.41(34)\\
\hline
\end{tabular}
\end{center}
\end{table}
Finally, we try to determine the position of the critical point. For this
purpose, we analyze $\chi_\infty$ and $\xi_\infty$ separately as 
\begin{eqnarray}
\log \chi_\infty(\beta) &=& A_\chi - \gamma_o \log (\beta_c - \beta), \nonumber \\
\log \xi_\infty(\beta)  &=& A_\xi -  \nu_o \log (\beta_c - \beta),
\label{fitchixi}
\end{eqnarray}
where $A_\chi$, $A_\xi$, $\gamma_o$, $\nu_o$, and $\beta_c$ are free parameters.
As before, we repeat the fit several times, including each time
only the data with $\beta\ge \beta_{\rm min}$.
Results using the infinite-volume data obtained with $\omega = 1$
are reported in Table~\ref{fit-omega1p0}. The estimates of 
$\beta_c$ show a tiny dependence 
on $\beta_{\rm min}$; moreover, the results obtained by using 
$\chi_\infty$ and $\xi_\infty$ are fully consistent. We take as
our final estimate 
\begin{equation}
 \beta_c = 1.08 \pm 0.01\pm 0.03,
\end{equation}
where the first error is statistical and the second one is systematic and 
takes into account the change in the estimate as $\omega$ varies 
between 0.6 and 1.4.
The data indicate therefore a transition point that is just beyond the 
interval of $\beta$ where we have made the simulations and thus they are 
compatible with the absence of a crossing point in $B_q$ and $\xi/L$ 
and with the fact that the estimates of these two quantities
at fixed $L$ get closer as $\beta$ increases. We can also perform a 
more quantitative check by using the results of \cite{KKY-06} for the 
critical-point values $B_q^*$ and $(\xi/L)^*$. They quote:
$B_q^* = 1.475(6)$ (bimodal distribution) and $B_q^* = 1.480(14)$
(Gaussian distribution); $(\xi/L)^*=0.627(4)$ (bimodal distribution)
and $(\xi/L)^*=0.635(9)$ (Gaussian distribution). These results are 
compatible with ours for $B_q$ and $\xi/L$ close to 
$\beta = 1.08$. For $\beta = 1.07$ we have 
$B_q = 1.411(4)$  ($L=12$), $B_q = 1.434(6)$ ($L=16$),  and
$\xi/L = 0.662(4)$ ($L=12$), $\xi/L = 0.648(4)$ ($L=16$).
These results are very close to the ISGM estimates and show the 
correct trend. They are therefore consistent with the conjecture that the 
ISGM and the SRAM belong to the same universality class.
Magnetic variables 
behave differently: as is clear in Fig.~\ref{Bim}, the estimates
at fixed $L$ are well separated, so that we do not expect any crossing at
$\beta\approx 1.08$. Thus, no criticality is expected in magnetic 
quantities: below the transition temperature the system is still
paramagnetic.

Using the results reported in Table~\ref{fit-omega1p0},
we can also estimate the critical exponents $\gamma_o$ and 
$\nu_o$. We obtain:
\begin{eqnarray}
&& \nu_o = 2.4 \pm 0.2 \pm 0.4 \\
&& \gamma_o = 5.3 \pm 0.3 \pm 1.0.
\end{eqnarray}
The estimates correspond to $\beta_{\rm min} = 0.88$. 
The first reported error is the statistical error and should 
be large enough to include most of the dependence on $\beta_{\rm min}$.
The second one gives the variation of the estimate with $\omega$.

\section{Conclusions}
\label{conclusions}

In this paper we study the SRAM by MC simulations
with the purpose of clarifying whether this model shows a critical 
behavior. We find clear evidence of FSS and criticality, with a 
critical point located at
\begin{equation}
   \beta_c = 1.08(4).
\end{equation}
Correspondingly, we compute the critical exponents associated
with the critical behavior of overlap observables. We obtain:
\begin{eqnarray}
   &&\eta_o = - 0.24(4) \\
   && \nu_o = 2.4(6) \\
   && \gamma_o = 5.3(1.3) .
\end{eqnarray}
These estimates are close to those obtained for the ISGM.
Ref.~\cite{Joerg-06} quotes $\nu_o = 2.22(15)$, $\eta_o = -0.349(18)$,
while \cite{KKY-06} reports 
$\nu_o = 2.39(5)$ and $\eta_o = -0.395(17)$. 
Other estimates are reported in Table 1 of \cite{KKY-06}.
With the quoted error bars there is a small discrepancy 
between our estimate of $\eta_o$ and those of 
\cite{Joerg-06,KKY-06}. This difference should not be 
taken too seriously, since there are similar discrepancies 
among the estimates obtained by different 
groups for the bimodal ISGM, see Table 1 of \cite{KKY-06}.
For instance, our result is close to $\eta_o = -0.26(4)$
reported in \cite{PC-99}. Note that \cite{Joerg-06}
also reports an estimate of the subleading exponent 
$\omega = 0.7(3)$. This result is obtained from fits of 
$\chi_\infty$ vs $\xi_\infty$. In this case, however,
analytic corrections are expected and therefore the leading 
scaling correction is $\xi_\infty^{-1/\nu_o} \approx 
\xi_\infty^{-0.45}$. Therefore, the result obtained 
in \cite{Joerg-06} provides only a strong indication 
that $\omega \gtrsim 1/\nu_o$ but does not give a quantitative
estimate of the exponent associated with the leading irrelevant 
operator. The bound $\omega \gtrsim 1/\nu_o$ is fully consistent 
with our results.

The possibility that the SRAM is in the same universality class
of the ISGM was put forward in \cite{CL-77} and found to be 
consistent with numerical results in two dimensions in \cite{BM-85}. It looks 
very plausible, since the SRAM is nothing but an Ising model
with local disorder and frustration. In some sense we can 
think of the SRAM and of the ISGM as two different versions of the 
same model: in the SRAM disorder is associated with lattice sites,
while in the ISGM disorder is associated with lattice bonds. 
They are analogous to the site-diluted and bond-diluted 
Ising model, whose Hamiltonian is given by Eq.~(\ref{raminfD})
with $j_{xy} = r_x r_y$ (site dilution) and 
$j_{xy} = r_{xy}$ (bond dilution), $r$ being a random variable such that 
$r = 1$ ($r=0$) with probability $p$ (resp. $1-p$). Note that 
the site-diluted model can also be seen as a bond-diluted model
with a correlated bond distribution, exactly as is the case for the 
SRAM. Nonetheless, there is little doubt---though a precise
numerical check is still missing---that the two models belong to 
the same universality class. 

Note that the SRAM is less frustrated 
than the standard ISGM, since in the 
SRAM $\overline{\prod_\square j_{xy}} = 1/27$. 
This fact does not rule out our conjecture 
since it is known that maximal frustration
is not necessary to obtain glassy behavior. For instance,
the random-bond Ising model with $j_{xy} = +1$ with probability $p$
and $j_{xy} = -1$ with probability $1-p$ has a glassy low-temperature
phase for \cite{OI-98} $0.233 \lesssim p \lesssim 0.767$.

It is interesting to generalize to SRAM by considering $N$-dimensional unit
vectors $\vec{u}_i$. The correlation of the bond variables around a lattice
plaquette becomes $\overline{\prod_\square j_{xy}} = 1/N^3$, which implies
that bond correlations vanish for $N\to \infty$. Thus, for $N = \infty$, the
SRAM is just an ISGM with a different continuous bond distribution.  In this
limit, therefore, the two models belong to the same universality class.  If
our conjecture for $N=3$ is correct, the same should hold for any $N\ge 3$.
For $N=1$ it is enough to redefine $\sigma_i \to u_i \sigma_i$ to reobtain the
standard ferromagnetic Ising model. The behavior for $N=2$ is unclear, since
this model is less frustrated than that with $N=3$ studied here. In
analogy with what happens in the random-bond Ising model mentioned in the
previous paragraph, it could have a glassy transition or a ferromagnetic
transition (the nature of the ferromagnetic transition is still object of
debate, see \cite{IOK-99}).

Our results are not precise enough to show conclusively that the SRAM and the
ISGM belong to the same universality class.  Because of the somewhat large
scaling corrections, one needs to perform simulations on larger lattices and
deeper in the critical regime (note that for $L=30$, our largest lattice, our
data extend only up to $\beta = 0.95$, that is quite far from the critical
point) to obtain precise and reliable estimates of the critical exponents.  In
particular, it would be interesting to see whether the small difference in the
estimates of $\eta$ disappears (at the same time it is important to include
scaling corrections in the analysis of the ISGM results, to understand the
reliability of the quoted error bars). Since we have reasonably precise
estimates for $G_4^*$ and $G_{22}^*$, it is also interesting to compute the
same quantities in the ISGM. This would provide an additional check of the
conjecture. Work in this direction is in progress.

A question that remains open is the behavior of the RAM for finite
anisotropy $D$. If there is indeed a low-temperature phase with
QLRO for small $D$ as predicted in \cite{AP-80,Feldman-00,Feldman-01},
then there should be a critical value $D^*$ such that ISGM behavior
is observed only for $D>D^*$. Nothing is known about $D^*$ 
and we cannot even exclude that $D^*=\infty$, so that ISGM behavior 
is observed only for model (\ref{raminfD}).


\section*{References}
\begin{thebibliography}{999}

\bibitem{HPZ-73}
Harris R, Plischke M and Zuckermann M J, 
{\em New model for amorphous magnetism},
1973 \PRL {\bf 31}, 160.

\bibitem{CPV-04}
Calabrese P, Pelissetto A and Vicari E,
{\em Spin models with random anisotropy and reflection symmetry},
2004 \PR E {\bf 70}, 036104 [cond-mat/0311576].

\bibitem{CHPRV-02}
Campostrini M, Hasenbusch M, Pelissetto A, Rossi P and 
Vicari E,
{\em Critical exponents and equation of state of 
the three-dimensional Heisenberg universality class},
2002 \PR  B {\bf 65}, 144520 [cond-mat/0110336].

\bibitem{PV-rev}
Pelissetto A and Vicari E, 
{\em Critical phenomena and renormalization-group theory},
2002 {\em Phys. Rep.} {\bf 368}, 549  [cond-mat/0012164].

\bibitem{DFH-05}
Dudka M, Folk R and Holovatch Yu, 
{\em Critical properties of random anisotropy magnets},
2005 {\em J. Magn. Magn. Mater.} {\bf 294}, 305 
[cond-mat/0406692].

\bibitem{IM-73}
Imry Y and Ma S-k,
{\em Random-field instability of the ordered 
state of continuous symmetry},
1975 \PRL {\bf 35}, 1399.

\bibitem{PPR-78}
Pelcovits R A, Pytte E and Rudnick J, 
{\em Spin-glass and ferromagnetic behavior 
induced by random uniaxial anisotropy},
1978 \PRL {\bf 40}, 476; {\em Erratum} 1982  {\bf 48}, 1297.

\bibitem{Aharony-75}
Aharony A, 
{\em Critical behavior of amorphous magnets}, 
1975 \PR B {\bf 12}, 1038.  

\bibitem{MG-82} 
Mukamel D and Grinstein G,
{\em Critical behavior of random systems},
1982 \PR B {\bf 25}, 381.

\bibitem{DFH-01a}
Dudka M, Folk R and Holovatch Yu,
{\em On the critical behaviour of random
anisotropy magnets},
2001 {\em Cond. Matt. Phys.} {\bf 4}, 77;
Dudka M, Folk R and Holovatch Yu, 2001
Phase Transition in the Random Anisotropy Model,
in {\em Fluctuating Paths and Fields}, 
edited by Janke W, Pelster A, Schmidt H-J and 
Bachmann M \/ (Singapore: World Scientific)
[cond-mat/0106334].

\bibitem{AP-80}
Aharony A and Pytte E, 
{\em Infinite susceptibility phase in random 
uniaxial anisotropy magnets},
1980 \PRL {\bf 45}, 1583;
{\em Low-temperature scaling for systems with 
random fields and anisotropies}
1983 \PR B {\bf 27}, 5872.

\bibitem{Feldman-00}
Feldman D E, 
{\em Quasi-long-range order in the random 
anisotropy Heisenberg model: 
Functional renormalization group in $4-\epsilon$
dimensions},
2000 \PR B {\bf 61}, 382. 

\bibitem{Feldman-01}
Feldman D E, 
{\em Quasi-long range order in glass states 
of impure liquid crystals, magnets, and superconductors},
2001 {\em Int. J. Mod. Phys.} B {\bf 15}, 2945 [cond-mat/0201243].

\bibitem{CL-77}
Chen J H and Lubensky T C,
{\em Mean field and $\epsilon$-expansion
of spin glasses},
1977 \PR B {\bf 16}, 2106.

\bibitem{MPV-87}
M\'ezard M, Parisi G and Virasoro M A, 1987
{\em Spin-Glass Theory and Beyond}\/ (Singapore: World Scientific).

\bibitem{FH-91}
Fischer K H and Hertz J A, 1991
{\em Spin Glasses}\/ (Cambridge: Cambridge University Press).

\bibitem{KR-03}
Kawashima N and Rieger H, 2004 
Recent progress in spin glasses in 
{\em Frustrated Spin Systems}, edited by Diep H T\/
(Singapore: World Scientific) [cond-mat/0312432].

\bibitem{JK-80}
Jayaprakash C and Kirkpatrick S,
{\em Random anisotropy models in the Ising limit},
1980 \PR B {\bf 21}, 4072.

\bibitem{Chakrabarti-87}
Chakrabarti A, 
{\em Spin-glass transition in three-dimensional 
random-anisotropy-axis model},
1987 \PR B {\bf 36}, 5747.

\bibitem{Fisch-90}
Fisch R, 
{\em Low-temperature behavior of random-anisotropy magnets},
1990 \PR B {\bf 42}, 540.

\bibitem{Fisch-98}
Fisch R, 
{\em Quasi-long-range order in random-anisotropy 
Heisenberg models},
1998 \PR B {\bf 58}, 5684 [cond-mat/9801033].

\bibitem{Itakura-03}
Itakura M, 
{\em Frozen quasi-long-range order 
in the random anisotropy Heisenberg magnet},
2003 \PR B {\bf 68}, 100405(R) [cond-mat/0303552].

\bibitem{HBDFFY-03}
Holovatch Yu, Blavats'ka V, Dudka M, von~Ferber C, Folk R and 
Yavors'kii T, 
{\em Weak quenched disorder and criticality: 
resummation of asymptotic(?) series},
2003 {\em Int. J. Mod. Phys.} B {\bf 16}, 4027
[cond-mat/0111158].

\bibitem{FHY-03}
Folk R, Holovatch Yu and Yavors'kii T,
{\em Critical exponents of a three dimensional 
weakly diluted quenched Ising model},
2003 {\em Physics Uspekhi} 46, 169 [cond-mat/0106468].

\bibitem{CEFPS-95}
Caracciolo S, Edwards R G, Ferreira S J, Pelissetto A
and Sokal A D, 
{\em Extrapolating Monte Carlo Simulations to 
Infinite Volume: Finite-Size Scaling at
$\xi/L \gg 1$},
1995 \PRL {\bf 74}, 2969 [hep-lat/9409004].

\bibitem{PC-99}
Palassini M and Caracciolo S,
{\em Universal finite-size scaling functions 
in the 3D Ising spin glass},
1999 \PRL {\bf 82}, 5128 [cond-mat/9904246].

\bibitem{CMPV-03}
Calabrese P, Mart\'\i n-Mayor V, Pelissetto A and Vicari E,
{\em Three-dimensional randomly dilute Ising model: 
Monte Carlo results},
2003 \PR E {\bf 68}, 036136 [cond-mat/0306272].

\bibitem{Joerg-06}
J\"org T, 
{\em Critical behavior of the three-dimensional 
bond-diluted Ising spin glass: Finite-size scaling 
functions and universality},
2006 cond-mat/0602215.

\bibitem{JLEM-06}
J\"org T, Lukic J, Marinari E and Martin O C, 
{\em Strong universality and algebraic scaling 
in two-dimensional Ising spin glasses},
2006 cond-mat/0601480.

\bibitem{KKY-06}
Katzgraber H, K\"orner M and Young A P,
{\em Detailed study of universality in 
three-dimensional Ising spin glasses},
2006 cond-mat/0602212.

\bibitem{BM-85}
Bray A J and Moore M A,
{\em Evidence for spin-glass behaviour in the 
random anisotropy-axis model}, 
1985 {\em J. Phys. } C: {\em Solid State} {\bf 18}, L139.

\bibitem{Houdayer-01}
Houdayer J,
{\em A cluster Monte Carlo algorithm for 2-dimensional
spin glasses},
2001 {\em Eur. Phys. J.} B {\bf 22}, 479 [cond-mat/0101116].

\bibitem{WS-05}
Wang J-S and Swendsen R H,
{\em Replica Monte Carlo simulation (Revisited)},
Proceedings of the conference 
{\em Statistical Physics of Disordered Systems and 
Its Applications}, Hayama (Japan), July 2004,
2005 {\em Progr. Theor. Phys. Suppl.} {\bf 157}, 317
[cond-mat/0407273].

\bibitem{Joerg-05}
J\"org T, 
{\em Cluster Monte Carlo algorithms for diluted spin
glasses}, 
Proceedings of the conference 
{\em Statistical Physics of Disordered Systems and 
Its Applications}, Hayama (Japan), July 2004,
2005 {\em Progr. Theor. Phys. Suppl.} {\bf 157}, 349 
[cond-mat/0410328].

\bibitem{Geyer-91}
Geyer C J, 1991 Markov chain Monte Carlo maximum likelihood,
in {\em Computer Science and Statistics:
Proc. of the 23rd Symposium on the Interface},
edited by E. M.~Keramidas \/
(Fairfax Station: Interface Foundation), p. 156.

\bibitem{GT-95}
Geyer C J and Thompson E A,
{\em Annealing Markov-chain Monte-Carlo with 
applications to ancestral inference},
1995 {\em J. Am. Statist. Association} {\bf 90}, 909.

\bibitem{HN-96}
Hukushima K and Nemoto K,
{\em Exchange Monte Carlo method and application 
to spin glass simulations},
1996 {\em J. Phys. Soc. Japan} 65, 1604
[cond-mat/9512035].

\bibitem{ED-05}
Earl D J and Deem M W, 
{\em Parallel tempering: theory, applications, 
and new perspectives},
2005 {\em Phys. Chem. Chem. Phys.} {\bf 7}, 3910 [physics/0508111].

\bibitem{KTHT-06}
Katzgraber H G, Trebst S, Huse D A and Troyer M, 
{\em Feedback-optimized parallel tempering Monte Carlo},
2006 {\em J. Stat. Mech.: Theory Expt.}, P03018 [cond-mat/0602085].

\bibitem{CEPS-95}
Caracciolo S, Edwards R G, Pelissetto A and Sokal A D,
{\em Asymptotic Scaling in the Two-Dimensional O(3)
$\sigma$ model at correlation length $10^5$},
1995 \PRL {\bf 75}, 1891  [hep-lat/9411009].

\bibitem{MPS-97}
Mana G, Pelissetto A and Sokal A D,
{\em Multigrid Monte Carlo simulation via XY embedding. 
II. Two-dimensional $SU(3)$ principal chiral model},
1997 \PR D {\bf 55}, 3674 [hep-lat/9610021].

\bibitem{CP-98}
Caracciolo S and Pelissetto A,
{\em Corrections to finite-size scaling in the lattice 
$N$-vector model for $N=\infty$},
1998 \PR D {\bf 58}, 105007 [hep-lat/9804001].

\bibitem{Privman}
Privman V (ed.), 1990 {\em Finite Size Scaling and 
Numerical Simulations of Statistical Systems}
(Singapore: World Scientific).

\bibitem{SS-99}
Salas J and Sokal A D, 
{\em Universal amplitude ratios in the critical 
two-dimensional Ising model on a torus},
2000 {\em J. Statist. Phys.} {\bf 98}, 551
[cond-mat/9904038].

\bibitem{Wegner-76}
Wegner F J, 1976 in {\em Phase Transitions and 
Critical Phenomena}, Vol. 6, edited by 
Domb C and Green M S\/
(New York: Academic) p. 7.

\bibitem{AF-82}
Aharony A and Fisher M E,
{\em Nonlinear scaling fields and corrections 
to scaling near criticality},
1982 \PR B {\bf 27}, 4394.

\bibitem{PV-98}
Pelissetto A and Vicari E, 
{\em Four-point renormalized coupling constant and 
Callan-Symanzik $\beta$-function in $O(N)$ models},
1998 {\em Nucl. Phys.} B {\bf 519}, 626
[cond-mat/9711078].

\bibitem{CCCPV-00}
Calabrese P, Caselle M, Celi A, Pelissetto A and Vicari E, 
{\em Non-analyticity of the Callan-Symanzik 
$\beta$-function of two-dimensional $O(N)$ models},
2000 \JPA {\bf 33}, 8155 [hep-th/0005254].

\bibitem{CHPV-02}
Caselle M, Hasenbusch M, Pelissetto A and Vicari E, 
{\em Irrelevant operators in the two-dimensional Ising model},
2002 \JPA {\bf 35}, 4861 [cond-mat/0106372].

\bibitem{GJ-87}
Guo H and Jasnow D, 
{\em Hyperuniversality and the renormalization 
group for finite systems},
1987 \PR B {\bf 35}, 1846.

\bibitem{CHPV-prep}
Campostrini M, Hasenbusch M, Pelissetto A and
Vicari E, 
{\em The critical exponents of the superfluid transition in 
$^4$He},
2006 cond-mat/0605083.

\bibitem{OI-98}
Ozeki Y and Ito N,
{\em Multicritical dynamics for the $\pm J$ Ising model},
1998 \JPA {\bf 31}, 5451.

\bibitem{IOK-99}
Ito N, Ozeki Y and Kikatani N,
{\em Non-universal critical behavior in the ferromagnetic transition 
of the $\pm J$ Ising model},
1999 {\em J. Phys. Soc. Jpn.} {\bf 68}, 803.

\end{thebibliography}
\end{document}